\documentclass[journal=jacsat,manuscript=article]{achemso}

\usepackage[version=3]{mhchem} 
\usepackage{bm,color}

\author{Ilya A. Eliseyev}
\affiliation{Ioffe  Institute, St.~Petersburg, 194021, Russia}
\author{Bogdan R. Borodin}
\affiliation{Ioffe  Institute, St.~Petersburg, 194021, Russia}
\author{Dmitrii  R. Kazanov}
\affiliation{Ioffe  Institute, St.~Petersburg, 194021, Russia}
\author{Alexander V. Poshakinskiy}
\affiliation{Ioffe  Institute, St.~Petersburg, 194021, Russia}
\author{Maja Rem\u{s}kar}
\affiliation{Jo\u{z}ef Stefan Institute, Jamova cesta 39, 1000 Ljubljana, Slovenia}
\affiliation{Ioffe  Institute, St.~Petersburg, 194021, Russia}
\author{Sergey I. Pavlov}
\affiliation{Ioffe  Institute, St.~Petersburg, 194021, Russia}
\author{Lyubov V. Kotova}
\affiliation{Ioffe  Institute, St.~Petersburg, 194021, Russia}
\author{Prokhor A. Alekseev}
\affiliation{Ioffe  Institute, St.~Petersburg, 194021, Russia}
\author{Alexey V. Platonov}
\affiliation{Ioffe  Institute, St.~Petersburg, 194021, Russia}
\author{Valery Yu. Davydov}
\affiliation{Ioffe  Institute, St.~Petersburg, 194021, Russia}
\author{Tatiana V. Shubina}
\affiliation{Ioffe  Institute, St.~Petersburg, 194021, Russia}

\email{ilya.eliseyev@mail.ioffe.ru}

\title[An \textsf{achemso} demo]
   {Twisted Nanotubes of Transition Metal Dichalcogenides with Split Optical Modes for Tunable Radiated Light Resonators}

\abbreviations{IR,NMR,UV}

\begin{document}

\begin{tocentry}

  \includegraphics[width=8.35cm]{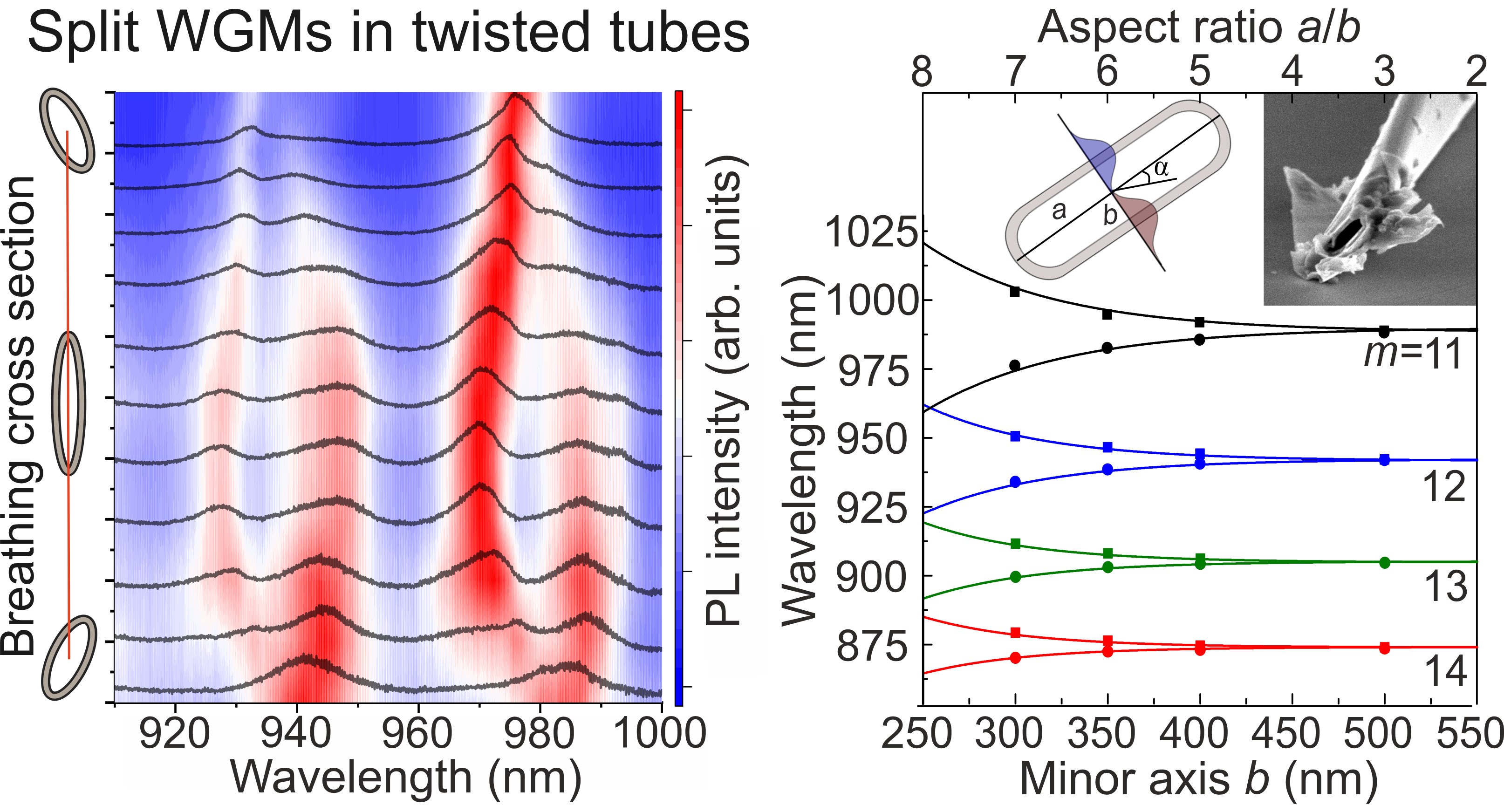}
Twisted Nanotubes of Transition Metal Dichalcogenides with Split Optical Modes for Tunable Radiated Light Resonators

\end{tocentry}

\begin{abstract}

Synthesized micro- and nanotubes composed of transition metal dichalcogenides (TMDCs) such as MoS$_2$ are promising for many applications in nanophotonics, because they combine the abilities to emit strong exciton luminescence and to act as whispering gallery microcavities even at room temperature. In addition to tubes in the form of hollow cylinders, there is an insufficiently-studied class of twisted tubes, the flattened cross section of which rotates along the tube axis. As  shown by theoretical analysis, in such nanotubes the interaction of electromagnetic waves excited at opposite sides of the cross section can cause splitting of the whispering gallery modes. By studying micro-photoluminescence spectra measured along individual MoS$_2$ tubes, it has been established that the splitting value, which controls the energies of the split modes, depends exponentially on the aspect ratio of the cross section, which varies in "breathing" tubes, while the relative intensity of the modes in a pair is determined by the angle of rotation of the cross section. These results open up the possibility of creating multifunctional tubular TMDC nanodevices that provide resonant amplification of self-emitting light at adjustable frequencies. 

\end{abstract}

The appearance of two-dimensional (2D) and tubular crystals in recent decades has become one of the most significant developments in the physics of nanostructures. These crystals are based on parent layered materials, primarily graphite and transition metal dichalcogenides. Carbon nanotubes (NTs) were first reported in 1991\cite{Iijima1991} followed by the demonstration of an extremely 2D carbon allotrope -- graphene\cite{Gaim2007}. In turn, the WS$_2$ and MoS$_2$ nanotubes were synthesized almost immediately after the carbon NTs in 1992 and 1993, respectively\cite{Tenne1992, Margulis1993}. The presentation of the MoS$_2$ monolayer as a direct-gap semiconductor was made in 2010\cite{mak2010atomically}, although the theoretical prediction of the direct-gap structure for a single-walled (\textit{i.e.} monolayer thick) MoS$_2$ nanotube was made much earlier, in 2000\cite{Seifert2000}. Thus, there is a certain similarity in the history of carbon and TMDC nanostructures, emphasizing the avant-garde role of nanotubes in this area of modern physics.

Unlike carbon NTs, the TMDC nanotubes are semiconducting at any folding and have a smaller band gap than in a bulk crystal\cite{Seifert2000, Seifert2000a}. Single-walled TMDC nanotubes are unstable\cite{Enyashin2007}, so their implementation  is a challenge\cite{Remskar2001, ghosh2020cathodoluminescence}. As a result, high-quality MoS$_2$ NTs synthesized by the chemical transport reaction, as well as other TMDC NTs fabricated by various methods,  are multi-walled crystals with the number of monolayers in the walls from a few to hundreds, and their diameter ranges from $\sim$30 nm to 2 or more microns\cite{Remskar1996,Remskar2021}. Recently, such individual MoS$_2$ NTs have been successfully used to create field-effect transistors\cite{Fathipour2015} and field emitters\cite{Lavrovski2020}, as well as to demonstrate quantum beats in the single electrons tunneling through a nanotube\cite{Reinhard2019}. Progress in the growth of MoS$_2$ nanotubes by sulphurizing the oxide nanowhiskers is significant\cite{Chithaiah2020}, and we can predict similar applications of them in the future. 

Exciton emission was discovered in TMDC nanotubes, both MoS$_2$ and WS$_2$, just a few years ago by several authors of this article using micro-photoluminescence spectroscopy ($\mu$-PL) \cite{kazanov2018multiwall, Shubina_tubes}. This opens the way for applications of such NTs in nanophotonics.
Since the Bohr radius of an exciton in MoS$_2$ is very small, on the order of a nanometer\cite{Yu2015}, the behavior of excitons in multi-walled NTs is generally similar to that in planar structures. In MoS$_2$ nanostructures, starting from the bilayer, indirect transitions have the lowest energy\cite{Cheiwchanchamnangij2012}, so they should be the main ones in optical processes. However, there is a certain competition between the recombination channels through direct ($\sim$1.8~eV) and indirect ($\sim$1.3~eV) excitons. Direct excitons have a giant oscillator strength and very fast radiative lifetime\cite{Wang2018, Palummo2015}, which ensures their sharp peaks in the absorption spectra of bulk material \cite{Evans1965} and the possibility of observing $\mu$-PL up to room temperature in multilayer flakes and NTs\cite{Shubina_tubes, Kotova2021}. Moreover, at low temperatures, both in nanotubes and in flakes, only direct exciton emission exists, while phonon-assisted indirect exciton radiation appears only at a temperature of 70–100 K\cite{Shubina_tubes, Smirnova2020}. This is a sign of a spin-forbidden (dark) exciton at the lowest state in the indirect exciton series\cite{Zhang2015}. A similar sequence of dark and bright exciton states has recently been observed for indirect exciton series in MoS$_2$ bilayers using a time-resolved temperature-dependent $\mu$-PL\cite{Eliseyev2021}. These data dictate the need to study indirect exciton radiation at room temperature when it can be applied in nanophotonics, which was done in this work.

A distinguishing feature of the $\mu$-PL spectra measured in individual NTs is the sharp peaks of whispering gallery modes (WGMs), which can appear in the regions of both direct and indirect excitons\cite{kazanov2018multiwall, Shubina_tubes}. Spectra with WGM peaks were modeled theoretically, assuming that the tubes are hollow cylindrical resonators with a wall thickness of tens of nanometers. Note that such modeling makes it possible to determine the number of monolayers in the tube walls and to distinguish a hollow tube from flakes and ribbons (collapsed NTs) in which the WGMs cannot exist. In addition, the existence of the WGM requires that an integer number of wavelengths inside coincide with the resonator circumference $m\lambda/n$ = $\pi D$, where $n$ is the effective refractive index, $\it {m}$ is the azimuthal angular number, and $D$ is the tube diameter. For a noticeable increase in radiation, a significant part of the electromagnetic energy must be concentrated inside the walls, which requires their certain thickness. This could be the reason why WGMs were not reliably detected in the optical spectra of TMDC NTs, which have thin walls and a small diameter, as in\cite{Sinha2021}.

The discovery of optical modes in TMDC nanotubes is of exceptional importance for their practical applications. The WGMs modify the radiation, making it polarized along the tube axis, which is useful for nanopolarizers\cite{kazanov2018multiwall}. They can accelerate slow indirect exciton transitions, making these structures suitable for room-temperature devices. For direct excitons, the strong light-matter interaction can promote the formation of exciton-polaritons in individual nanotubes\cite{kazanov2020towards}. These points correspond to the general trends in the development of nanophotonics and optoelectronics based on TMDCs. For example, atomically thin TMDCs have been integrated with a photonic crystal to control the rate of spontaneous emission\cite{gan2013controlling}, while freestanding microdisks \cite{ye2015monolayer} and Bragg cavities\cite{shang2017room} were used to create efficient lasers. A successful attempt was made to form optical micro- and nanocavities from bulk TMDCs themselves\cite{li2014measurement}. Undoubtedly, the best solution in this area is the use of micro- and nanotubes as natural optical resonators.

Until now, the main attention has been paid to micro- and nano- MoS$_2$ tubes with a circular cross-section.  These rigid crystals are grown in the helical mode from the vapor phase under almost equilibrium conditions. In addition, there is a class of tubular structures whose general geometry is very different from a hollow cylinder\cite{remvskar1998mos}. They appear due to a local change in the growth conditions, which causes an imbalance in the total energy, controlled by the strain and elasticity of the bonds \cite{Enyashin2007}. It was assumed that the flattening of the cross section is caused by the deviatoric stress along the perimeter of the section when it exceeds a certain threshold \cite{Kralj2002}. With strong deformation, when van der Waals forces begin to act between the opposite walls of the tube, a complete collapse into a continuous ribbon is possible\cite{Chopra1995}. The modified form corresponds to the absolute minimum of elastic energy. One can assume that changing the shape of nanotubes opens up a way to selectively control the intensity, frequency, or direction of radiation, as was done in modified WGM disk resonators\cite{reed2015wavelength, jiang2016whispering}.

In this work, we extend the class of light-emitting TMDC micro- and nanotubes with twisted tubes possessing a flattened cross section, which, nevertheless, are capable of supporting optical modes similar to whispering gallery modes in cylindrical tubes. Using $\mu$-PL spectroscopy of individual tubes, we demonstrate their distinctive feature, the splitting of WGMs when electromagnetic fields circulating on the opposite  tube walls can interact \textit{via} a narrow gap in a highly flattened cross section. A model is presented that makes it possible to describe the spectra in twisted tubes of two types: with a constant cross section and with a cross section that varies along the tube axis ("breathing" tubes). Both types demonstrate an antiphase variation in the intensity of a pair of split modes as the cross section rotates, while a change in cross section shape along the axis of the breathing tube ensures a gradual change in the splitting energy. Our research creates a tunable platform for resonant luminescence amplification in TMDC-based tubular nanophotonic devices.

\section{Split WGM doublets in flattened nanotubes}

First, we present the theoretical classification of the optical WGMs in NTs with flattened cross sections.  The first approximation is to consider the wall of the nanotube as a slab resonator with periodic boundary conditions.  Then, we obtain the set of doubly degenerate modes with the frequencies  $\omega_m^{(0)} = 2\pi c m/(n_{\rm eff}L)$, where $L$ is the cross section circumference and  $n_{\rm eff}$ is the effective refractive index of the slab resonator mode. In the NT with a circular cross section, this degeneracy originates from the $C_\infty$ rotational axis that can transform one mode into the other.  However, in the NTs with deformed cross section, the degeneracy must be lifted, and the splitting of WGMs into doublets shall arise.

\begin{figure}[h]
\centering
  \includegraphics[width=0.6\linewidth]{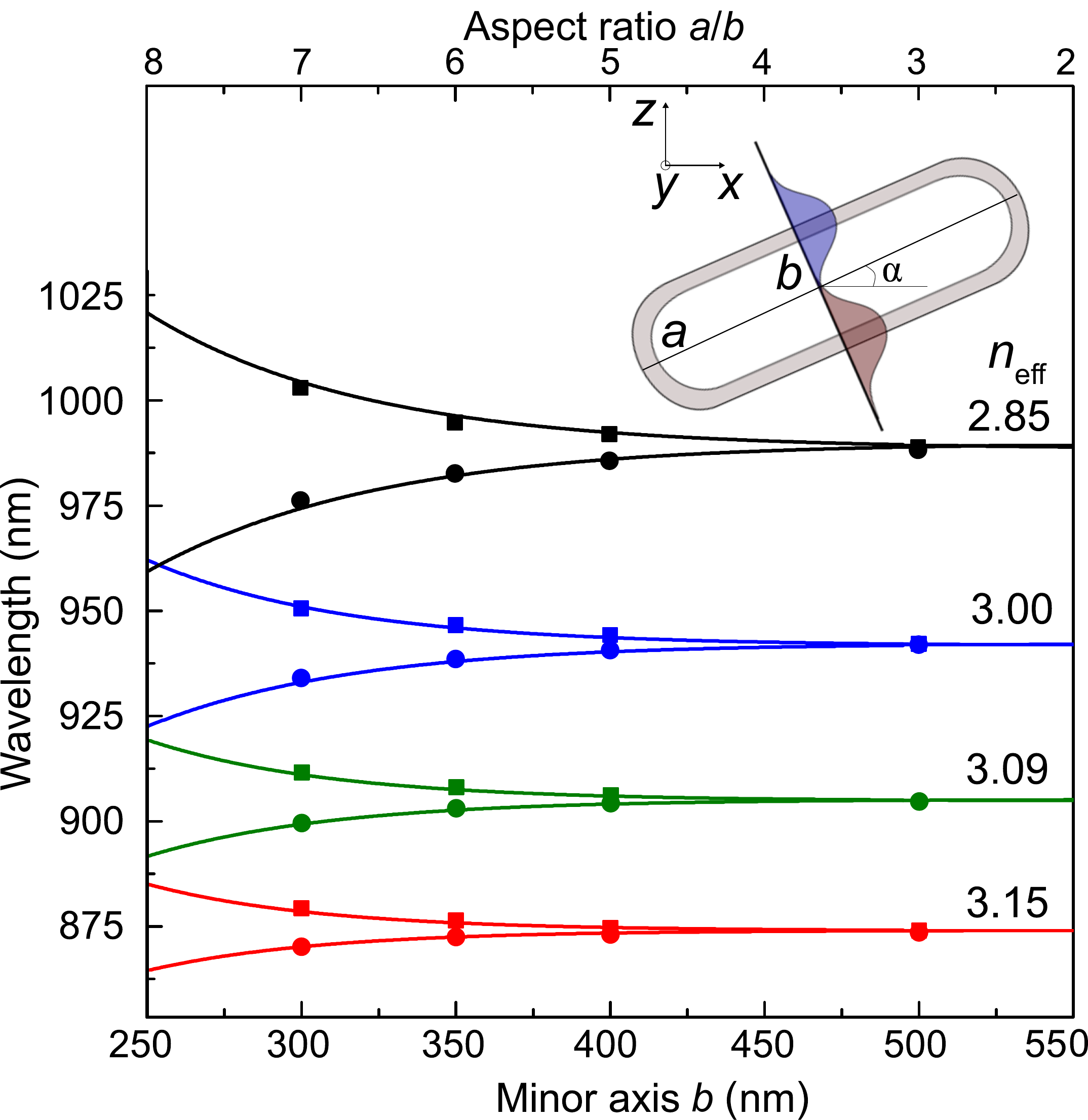}
  \caption{Splitting of the WGMs with different $m$ in nanotubes with the fixed circumference $L=3.5~\mu$m and different size of the minor axis $b$; the aspect ratio is shown on the upper plot axis. Numbers above the lines indicate the effective refractive index $n_{\rm eff}$ at the energies of studied optical modes. The inset shows the distribution of mode energy at opposite sides of a nanotube.}
  \label{fig:theory}
\end{figure}

To demonstrate the origin of the splitting, we consider a NT with the cross section of the racetrack shape, characterized by the two axes: major $a$ and minor $b$, and a constant wall width $d \ll a,b$, see geometry in inset in Figure~\ref{fig:theory}. At $a=b$, the model reduces to the 
case of the circular nanotube. At $a>b$, the tails of the electromagnetic field in the two long sides of the racetrack overlap leading to the coupling, see the inset in Figure~\ref{fig:theory}. The eigenmodes are then the symmetric and anti-symmetric superposition of the fields in the two sides of the racetrack with the energies
\begin{equation}\label{eq:split}
    \omega_m^{(e,o)} = \omega_m^{(0)} (1  \mp t {\rm e}^{-\kappa b}) \,.
\end{equation}
Here, indices ``e'' and ``o'' correspond to modes that are even and odd with respect to the inversion along the nanotube major axis, $\kappa = ({\omega}/{c})\sqrt{n_{\rm eff}^2-1}$ is the spatial decay rate of the field tails, and $t$ is a dimensionless parameter. 

Figure~\ref{fig:theory} shows the dependence of the WGM energies on the length of the nanotube cross section minor axis $b$, obtained from the rigorous numerical solution of Maxwell equations. One can see that WGMs split as $b$ decreases. The splitting value is well described by Eq.~\eqref{eq:split} with $t \approx 1$. Also note that the splitting is larger for the modes with lower $m$ due to smaller value of $\kappa \propto \omega$, see Eq.~\eqref{eq:split}. 

Now, we describe the optical selection rules for excitation and detection of the split WGM doublet. 
We focus on the modes that are polarized along the nanotube axis. The efficiency of the mode coupling to the free electromagnetic wave outside the NT is determined by the overlap
\begin{align}\label{eq:Am}
   A_m^{(e,o)} = \int E_m^{(e,o)}(\bm r)\, {\rm e}^{{\rm i} \bm k \cdot \bm r} \, d\bm r \,,
\end{align}
where $E_m^{(e,o)}(\bm r)$ are the electric field distributions of the even and odd WGMs in the $m$-th doublet, $\bm k$ is the wave vector of the light outside the NT. 

The overlap $A_m^{(e,o)} $ depends on the angle between the minor axis of the NT cross-section and free light wave vector, which is supposed to be perpendicular to the NT axis.
To investigate this dependence, it is instructive to consider the case of a weakly deformed nanotube. Then, the WGMs are close to that of the circular nanotube and have the azimuthal dependence $E_m^{(e)} \sim \cos m (\varphi-\alpha)$ and $E_m^{(o)} \sim \sin m (\varphi-\alpha)$, where $\alpha$ is the angle between the $y$ axis and the reflection symmetry axis of the cross section, see inset in Figure~\ref{fig:theory}. Then, the efficience of mode coupling to the light propagating in $z$ direction calculated from Eq.~\eqref{eq:Am} reads
\begin{align}\label{eq:Intensity2}
|A_m^{(e)}|^2 \sim J^2_m (R\omega/c) \,{\cos}^2 \left[ m \left(\alpha+\frac{\pi}{2}\right) \right] \,,
\\
|A_m^{(o)}|^2 \sim J^2_m (R\omega/c) \,{\sin}^2 \left[ m \left(\alpha+\frac{\pi}{2}\right) \right]\,.
\end{align}
Note that total efficiency for the two split modes $|A_m^{(e)}|^2+|A_m^{(o)}|^2$ does not depend on angle $\alpha$. The main conclusion is that the intensities of the split WGMs are very sensitive to the nanotube orientation. They change between minimum and maximum under rotation by the angle $\Delta \alpha = \pi/(2m)$.

Thus, in the flattened NTs, splitting of the WGM peaks is anticipated. The degree of splitting depends on the aspect ratio of the cross-section and the mode number. The relative intensity of the split odd and even modes is controlled by the angle between the minor axis and the direction light incidence. 

\section{Twisted MoS$_2$ tubes: Synthesis and characterization}

To verify the theoretical predictions, we studied  MoS$_2$ micro- and nanotubes. 
The nanotubes under study were obtained by a chemical vapor transport reaction using iodine as a transport agent \cite{Remskar1996, Remskar2011, Remskar2021}. They are usually synthesized together with thin and curved MoS$_2$ flakes. In fact, tube nucleation occurs in the folds or curved edges of these flakes, where the growth of the next basal MoS$_2$ plane cannot follow a planar structure. Continued growth can be in the form of scrolls, which can transform into tubes if one of the preferred stacking orders (2\textit{H}, 3\textit{R}) is performed and van der Waals forces begin to act between the monolayers. Furthermore, the NTs grow in a chiral mode with a similar angle of chirality in different tubes. This chiral growth can be a possible reason for the appearance of twisted NTs. A demand for satisfying one of two possible stackings gets very challenging with increasing a tube diameter due to the increase in the number of layers.  Lattice deformation can cause helical twisting which tends to collapse the tube, but if the number of layers is high enough, the elastic energy will prevent the total collapse. Most of the previous research have been devoted to nanotubes in the form of hollow cylinders. However, careful examination showed that there are nanotubes with a deformed cross section.

\begin{figure}[H]
\includegraphics[width=1\textwidth]{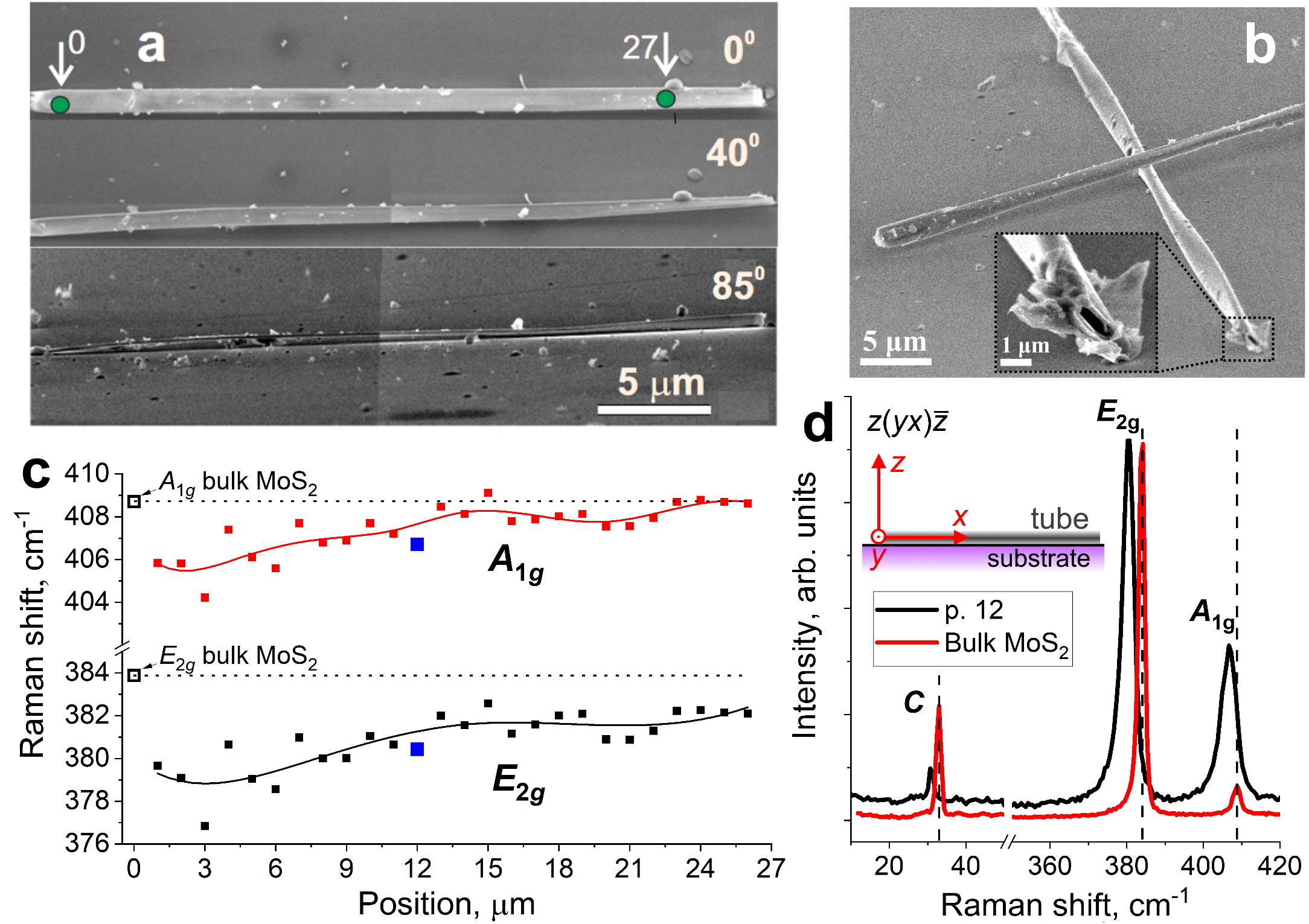}
\caption{\label{fig:SEM_Raman} \textbf{(a) SEM images of a MoS$_2$ nanotube from different viewing angles. (b) Nanotube tail showing a void in its racetrack-like cross section. (c) Change in the position of the $E_{2g}$ and $A_{1g}$ Raman modes along the axis of the tube shown in (a). The data were obtained from Raman spectra measured at room temperature in the $z(yx)\Bar{z}$ geometry. The polarization geometry is given in Porto's notation, where the $z$ axis is orthogonal to the substrate plane, and $x$ coincides with the tube axis, as it is shown on the inset. (d) Comparison of the Raman spectra of the nanotube measured at the point marked in Figure~\ref{fig:SEM_Raman}c by a blue square (black curve) and of a bulk MoS$_2$ flake (red curve).}} 
\end{figure}

In our study, individual tubes were isolated from the mixture with tweezers and placed on a Si/SiO$_2$ substrate with subsequent precise positioning using an atomic force microscope. The detailed procedure of positioning is described in Supporting Information (SI) Figure S1 and accompanying text. The shape of the nanotubes, according to scanning electron microscopy (SEM) (Figure~\ref{fig:SEM_Raman}a, b), is flattened and twisted with the racetrack or highly elliptical cross section. A series of SEM images at increasing angles of view of the nanotube shown in Figure~\ref{fig:SEM_Raman}a reveal that the cross section of the tube is flattened with characteristic dimensions of approximately 1.1 $\mu$m, and 200 nm (for details, see SI, Figures S2 and S3). Such a shape was previously reported in \cite{remvskar1998mos} and is connected to the peculiarities of the tube growth process. Figure~\ref{fig:SEM_Raman}b shows a pair of nanotubes placed one above another to form a cross (optical image of the NTs presented in SI, Figure S4). The upper tube exhibits a non-monotonic change in diameter. This indicates a "breathing" shape of the tube. The bottom tube has a pronounced twisting along the growth axis. Inset in Figure~\ref{fig:SEM_Raman}b also shows a flattened cross section of the bottom NT. 

Now we discuss mechanisms defining the NT shape. The nanotube begins to grow with a cylindrical shape, and further growth leads to an increase in the degree of ellipticity of the tube's cross section with rotation of the long axis of the ellipsoid along the nanotube axis. In the extreme case, a nanotube is collapsed to the ribbon\cite{Chopra1995}. We assume that the shape of the nanotube is determined by a balance among elastic force, van-der-Waals force, and internal tensile strain accumulating during the growth process\cite{Kralj2002,Enyashin2007}. In the case of thin-wall nanotubes, the van-der-Waals force might be sufficient to overcome the elastic force and glue nanotube walls together, forming a nanoribbon. In the case of thick-wall microtubes, the balance is the opposite. The elastic force of thick walls is much greater than the van-der-Waals force, and the tube tends to have an elliptical cross-section. According to the model presented in the work\cite{Chopra1995}, in our case, the walls of the tubes are several tens of nanometers thick, and the elastic force significantly exceeds the van-der-Waals force. 

Therefore, we believe that source of the observed shape is internal tensile strain. Indeed, the typical Raman spectrum of the nanotube (Figure~\ref{fig:SEM_Raman}c,d) shows that it consists predominantly of 2\textit{H} layers with a tensile strain. The first claim is supported by the presence of a low-frequency shear mode at 31 cm$^{-1}$, which should be absent in 3\textit{R}-MoS$_2$ with $N>3$ \cite{van2019stacking}. The presence of the tensile strain is confirmed by Raman data in Figure~\ref{fig:SEM_Raman}c,d. It can be seen that the $E_{2g}$ and $A_{1g}$ lines are redshifted with respect to their position in bulk MoS$_2$. The magnitude of the shift, and hence the strain, generally tends to increase as one moves along the growth axis (Figure~\ref{fig:SEM_Raman}c); however, as we move along the tube, we see that the magnitude of strain slightly oscillates, which is a sign of periodic relaxation and increase in strain. The observed behavior is typical for the studied nanotubes. The tensile strain was also observed in TMDC tubes studied by other groups \cite{staiger2012excitonic,ghosh2020cathodoluminescence}. The reason for the observed character of strain is layer-by-layer growth and different curvature of the neighboring layers in the radial direction. 

Starting part of the nanotube has a higher number of layers and a cylindrical shape. Deformation of the cross section of the nanotube at the beginning of the growth axis is difficult due to a large number of layers and the high stiffness of the walls. While we move in the axial direction, the thickness of the walls is reduced, but the layers already have the tension strain. Due to the inverse Poynting effect, this should lead to torsional deformation of the tube\cite{goldstein2018chiral} or twisting.

Thus, during the growth, the emergence of the tensile strain leads to the twisting of the nanotubes. Along the growth axis, a variation of the tensile strain was revealed by Raman spectroscopy that may indicate a formation of the breathing NT, in which the degree of flattening of the cross section is varied. It is worthy to note that twisting was also observed for the nanotubes with a smaller diameter of 100-200 nm and was revealed by transmission electron microscopy (see SI, Figure S5). 
In our study of optical modes in individual twisted nanotubes, we concentrated on two types of nanotubes - breathing ones (see the upper tube in Figure~\ref{fig:SEM_Raman}b) and with a rotating flattened cross section of constant shape (Figure~\ref{fig:SEM_Raman}a).

\subsection{$\mu$-PL and optical modes in twisted tubes}

\begin{figure}[t!]
\includegraphics[width=1\textwidth]{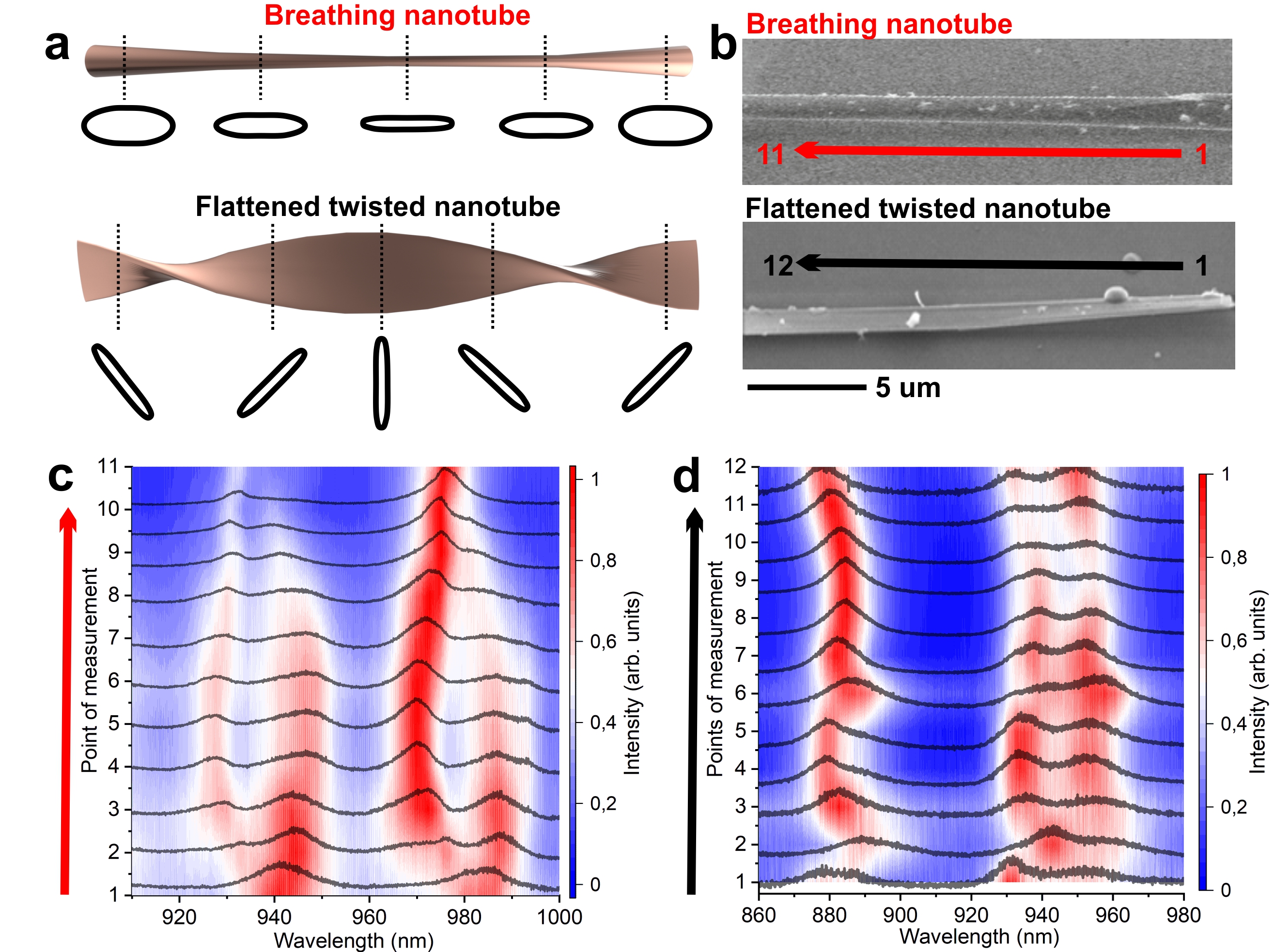}
\caption{\label{fig:PL_exp} \textbf{(a) Illustration of two tube types and their cross-sections. (b) SEM images of the measured tubes: 1 — breathing tube with a variable degree of flattening, 2 — flattened twisted tube with a stable flattened cross section. (c, d) $\mu-PL$ spectra measured at room temperature in tube 1 (c) and in tube 2 (d). All the spectra were measured in the $z(yx)\bar{z}$ polarization, the axes orientation is shown on Figure \ref{fig:SEM_Raman}d.}}
\label{mPL_tubes1_2}
\end{figure}

As it is clear from the characterization, we deal with very peculiar objects. TMD tubes with flattened cross-sections and thick ($\geq$50 nm) walls can accommodate whispering gallery modes that determine their optical properties\cite{Shubina_tubes,kazanov2020towards}. Here, we consider the flattened tubes with a smooth variation of flattening degree and rotation of the cross-section along the tube, which creates conditions for observing modified WGM resonances. To provide a conclusive analysis of optical properties, it is necessary to investigate the PL spectra considering the shape and cross section flattening degree at each measurement point. Such an investigation is presented in Figure \ref{fig:PL_exp}. Figure \ref{fig:PL_exp}a illustrates the shape of two types of tubes investigated and their cross-sections constructed based on SEM images from various angles. The "breathing" tube (upper NT in Figure~\ref{fig:SEM_Raman}b) has a circumference $L=3.5~\mu$m and a race-track cross-section with a degree of flattening that varies along the length. In the middle of the measured segment, the tube is severely flattened (aspect ratio $\approx$ 7:1), while on the ends, it is significantly more circular (aspect ratio $\approx$ 4:1). The "flattened twisted" tube (Figure~\ref{fig:SEM_Raman}a) has a circumference $L=2.5~\mu$m and a constant, strongly flattened (aspect ratio $\approx$ 6:1) race-track cross-section and rotates along its long axis. 

Figure \ref{fig:PL_exp}b shows a SEM image of the two tubes with indicated measurement sections. The $\mu$-PL technique used provides a measurement step of 1 $\mu$m. We scrupulously investigated the spectra of both tubes. The spectra are demonstrated in Figure \ref{fig:PL_exp}c, d. Figure \ref{fig:PL_exp}c demonstrates spectra obtained from the breathing tube. The first spectrum revealed four peaks (the most intense two of them are shown in the graph, full spectra are presented in SI, Figure S6). Positions of the peaks correspond to the indirect exciton in MoS$_2$ NT\cite{Shubina_tubes}. PL intensity of the direct exciton  in ~700 nm spectral range is suppressed (see SI, Figure S6). Observation of the two peaks in the 900-1000 nm spectral range was expected for the case of slightly flattened ends. These peaks are specific to circular tubes of this size and were observed in the previous works\cite{Shubina_tubes,kazanov2020towards}. The origin of their manifestation is the whispering gallery modes within the tube walls. Moving the measurements along the tube, we can see that each original peak smoothly splits into two. The maximum splitting ($\approx$ 20 nm) is observed in the middle of the examined section (see spectrum 6). It should be noted that the maximum degree of flattening is also observed at this point. Thus, we can assume that the peak splitting is induced by the interaction between whispering gallery modes from the top and bottom flat walls. Then, moving forward to the expanding part of the tube, we can see that the split peaks begin to merge. At point 11, the tube is also slightly flattened, as it was at the beginning, and we can see the two original peaks. Such an evolution of the positions of the split peaks is in an excellent agreement with the theoretical predictions in Figure~\ref{fig:theory}.  

Figure \ref{fig:PL_exp}d demonstrates spectra obtained from the flattened twisted tube. SEM images of the tube revealed a constant degree of flattening, and we do not observe any pronounced evolution of splitting along the tube length. However, one can see that in the first third of the measured section (spectra 1-5), the left of two split peaks near 950 nm is more pronounced. Then (curves 6-7), the right peak becomes dominant. Next (spectra 8-9), the left prevails again, and then (spectra 11-12), the right. Such behavior of intensity is correlated with the period of twisting. Thus, it can be assumed that the angle between the incident light and the tube plane can also affect the efficiency of excitation of a particular mode in the tube.

\section{Discussion}

The model of the twisted NT with the flattened cross section, described in the previous sections, can be used to simulate the PL spectra. We suppose that both major and minor axes, $a$ and $b$, change along the nanotube, but the circumference $L = \pi b + 2(a-b)$ remains constant. Additionally, the nanotube cross-section rotates with respect to the PL detection direction. 

To model the PL spectra, we use the approach developed in Refs.~\cite{kazanov2018multiwall, kazanov2020towards}. 
First, the scattering problem for the incident pump wave is solved numerically, and the resulting electric field distribution $\bm E_i(\bm r)$ in studied polarization is calculated. Note that the problem can also be solved analytically by decomposing the incident plane wave over angular harmonics, $e^{{\rm i} k_z r \rm{sin} (\varphi-\alpha)}=\sum\limits_{n=-\infty}^{+\infty} J_n (k_z r) e^{{\rm i} n (\varphi-\alpha)}$, and considering their scattering separately. From the calculated pump field distribution, the local PL generation rate is obtained as $\propto |\bm E_i(\bm r)|^2$. We also assume that polarization is lost during the relaxation process.
To calculate the PL emission in a certain direction, we use the Lorentz reciprocity. The time-reversed problem, where the light is incident from that direction, is solved and the field distribution $\bm E_f(\bm r)$ is calculated. The emission intensity is then determined by 
$I \propto \int |E_f(\bm r)|^{2} |E_i(\bm r)|^2 d\bm r$. 

\begin{figure}[h]
\centering
  \includegraphics[width=0.6\linewidth]{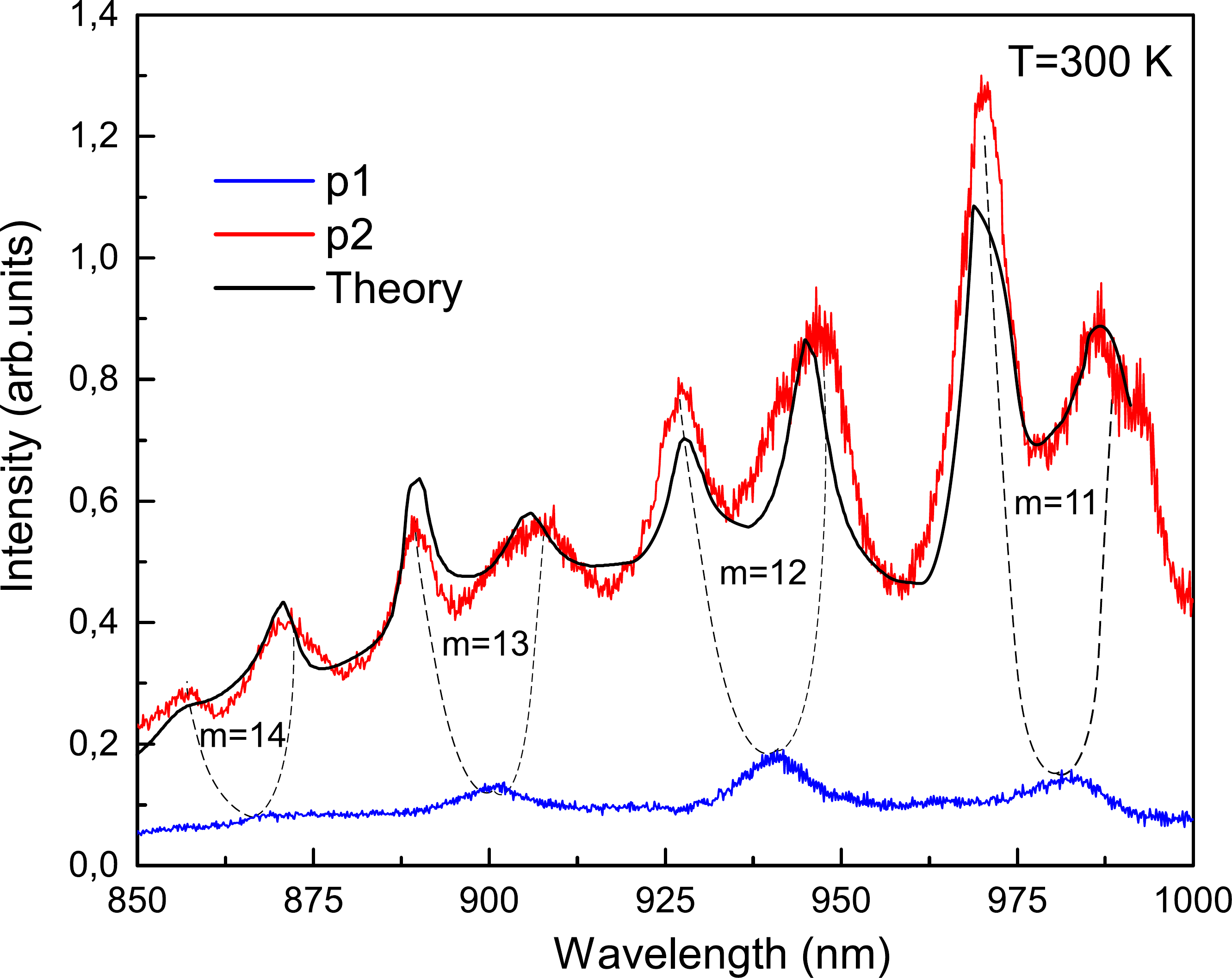}
  \caption{\textbf{Modeling of the experimental $\mu$-PL spectra detected in the region of indirect exciton emission in the breathing nanotube.  The simulation is made using the developed model for a twisted NT with a flattened cross-section (black line). Red and blue lines show the measured spectra in different points (spectrum 6 and 1, respectively, from Figure~\ref{fig:PL_exp}c}).
   }
  \label{fig:experiment_theory}
\end{figure} 

Figure~\ref{fig:experiment_theory} shows the calculated PL spectrum for the polarization along the nanotube axis ($x$) together with the experimental spectra from two particular points corresponding to the weakly (spectrum 1 in Figure~\ref{fig:PL_exp}c, blue line) and strongly (spectrum 6 in Figure~\ref{fig:PL_exp}c, red line) flattened cross section of the breathing tube. 
Shown in Figure~\ref{fig:experiment_theory} by the black line is the calculated PL spectrum for the considerably flattened breathing NT with the circumference $L = 3.5\,\mu$m, $a/b = 7$ and wall width of 70 nm. Additionally, the cross section was assumed rotated by the angle $\alpha =8^{\circ}$ in order to reproduce the correct relative intensities of the WGM peaks, see discussion below.

The spectra of the flattened twisted NT in Figure~\ref{mPL_tubes1_2}d shows peaks at frequencies that are shifted compared to the spectra of the breathing NT due to different geometrical characteristics. The circumference of this NT is $L = 2.5\,\mu$m, $a/b = 6$ with the same wall width of 70 nm. Consequently, the inter-mode distance increases along with the gap between the split peaks originating from previously  degenerate modes, so the PL in the selected spectral region has a pair of split modes at $\sim$ 880 nm and $\sim$ 940 nm with an angular momentum $m=9$ and the left peak of the neighboring lower energy pair with $m=8$. 
Also, the peaks do not "split and join" along the tube axis from points 1 to 12, as in the case with breathing NTs, since this twisted NT has a nearly constant cross-sectional shape with a stable gap between opposite walls.

\begin{figure}[t]
\centering
  \includegraphics[width=0.99\linewidth]{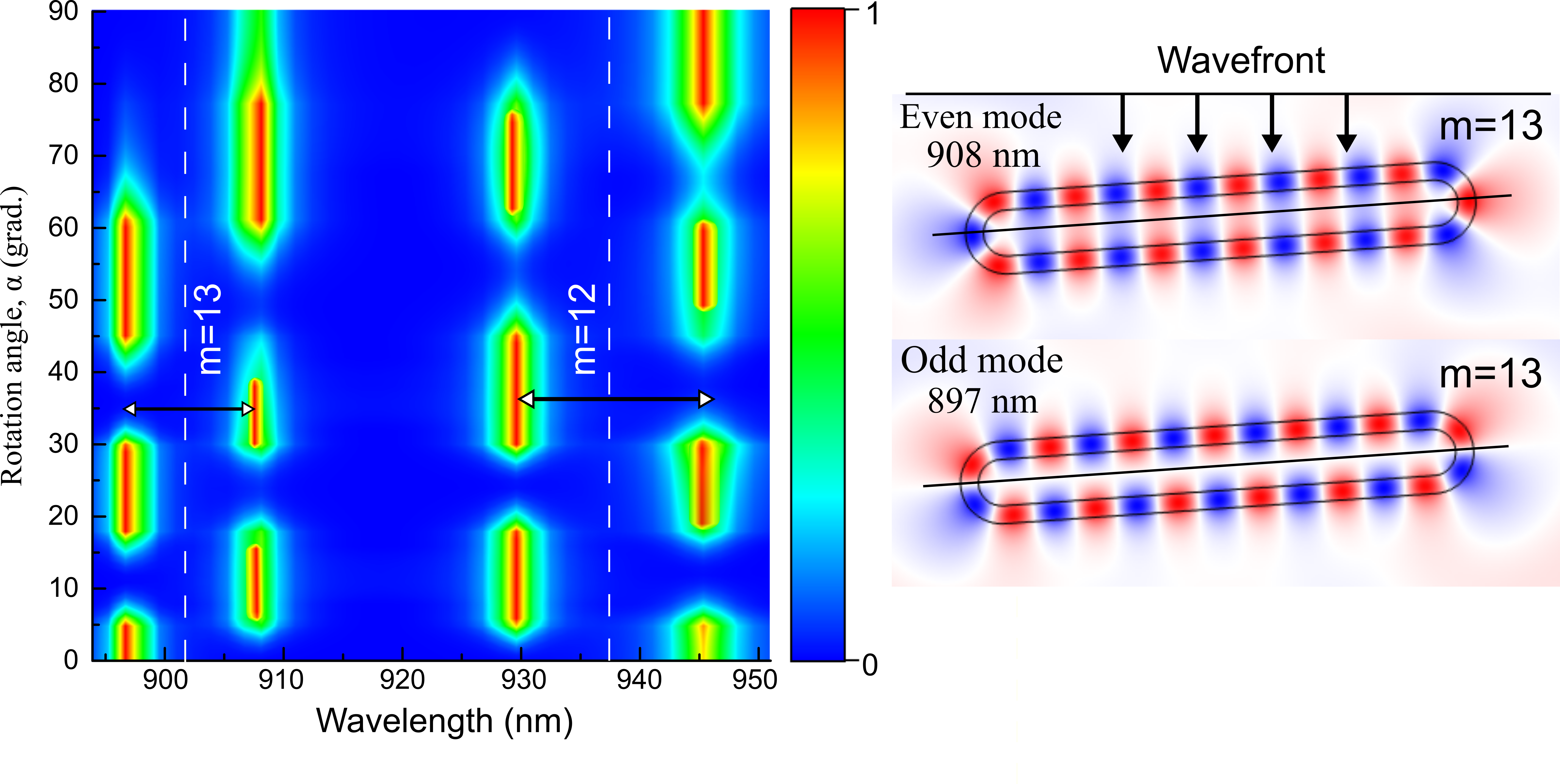}
  \caption{\textbf{Calculated PL intensity as a function of  the angle of cross section rotation $\alpha$ with respect to the detection direction. The considered spectral range contains two WGMs with angular numbers $m=12$ and $m=13$, each of which is split into even and odd modes. The right panel shows the distribution of the electric field for the split pair with $m=13$. The calculations were carried out using the parameters of the breathing NT at the point with the maximum mode splitting ($L = 3.5\,\mu$m, $a/b=7$) assuming a racetrack cross section.}
  }
  \label{fig:PL_angle1}
  \end{figure} 

\begin{figure}[t]
\centering
  \includegraphics[width=0.99\linewidth]{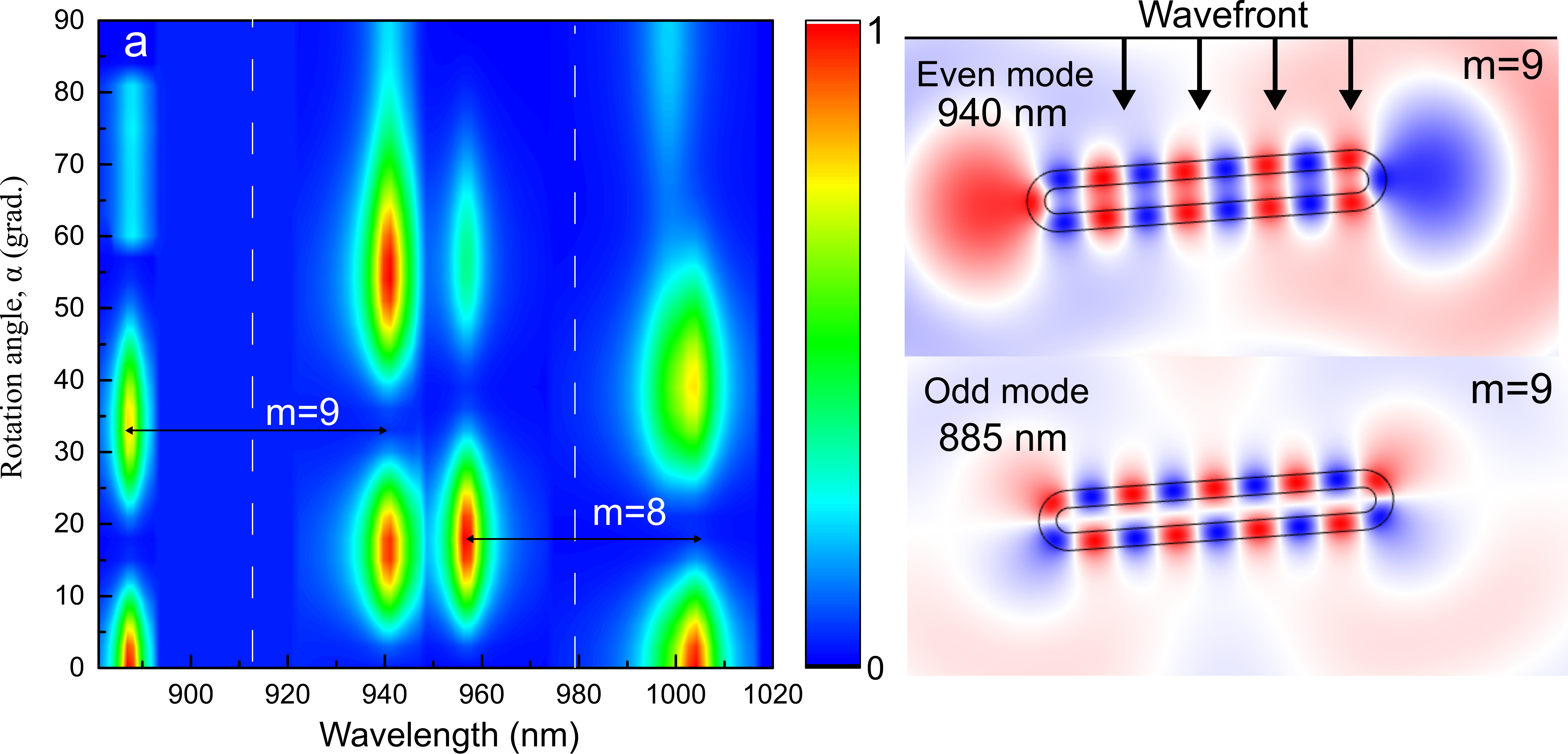}
  \caption{\textbf{Calculated PL intensity as a function of the angle of cross section rotation $\alpha$ with respect to the detection direction, obtained for the flattened twisted nanotube ($L = 2.5\,\mu$m, $a/b=6$, racetrack cross section) in the spectral range near two WGMs with angular numbers $m=8$ and $m=9$. The right panel shows the distribution of the electric field for a split pair with $m=9$}. 
  }
   \label{fig:PL_angle2}
  \end{figure} 
  
Now we investigate in more detail the dependence of the WGM peak intensities on the orientation of the twisted nanotube cross section with respect to the excitation and detection axis. For the sake of simplicity, we assume that the excitation is off-resonance, so that $|E_i(\bm r)|^2$ is distributed uniformly along the cross section circumference. In contrast, the detection is performed at the frequency $\omega$ that is close to the frequency $\omega_m^{(e,o)}$ of a particular WGM. 
Then, the intensity of the peaks in the PL spectrum is proportional to $|A_m^{(e,o)} |^2$, where $A_m^{(e,o)}$ is given by Eq.~\eqref{eq:Intensity2}. 

Figure~\ref{fig:PL_angle1} shows the variation of the PL spectra on the angle $\alpha$ of cross section rotation with respect to the detection direction, calculated using the parameters of the breathing NT in the point of maximal mode splitting. Here, to demonstrate the effect, the angle $\alpha$ changes from 0$^\circ$ to 90$^\circ$.
Spectrum modelling shows the presence of WGMs with angular numbers $m=12$ and $m=13$, splitting due to cross-sectional deformation to the shape of a racetrack with $a/b \sim 6$. 
Importantly, the intensities of pairs of split modes with the same $m$ oscillate with $\alpha$ in opposite phases, as it expected from the equation~\eqref{eq:Intensity2}.
The number of oscillations in the range $0 \leq \alpha \leq 90^\circ$ is $m/4$.

Similar calculations were performed using the parameters of the flattened twisted nanotube, the PL spectra of which are shown in Figure \ref{fig:PL_exp}d. According to the simulation, they contain WGMs with the angular number $m=8$ and $m=9$ (Figure~\ref{fig:PL_angle2}).  For neighboring pairs of split WGMs, the intensities of the even and odd modes are also swapped, as follows from Eq.~\eqref{eq:Intensity2}. 
Due to the smaller circumference $L = 2.5\,\mu$m than in the previously discussed NT ($L = 3.5\,\mu$m), both the free space region between WGMs and the mode splitting value are increased. As a result, the intense spot of the odd component with $m=8$ occurs close to that of the even component with $m=9$. In the general case, this fact, which is characteristic of strongly flattened nanotubes of a small size, should be taken into account when analyzing optical modes in their PL spectra.

The alternating pattern of the split WGM intensities is clearly seen in the experimental PL spectra. For example, in the spectrum of a breathing NT (see Fig.~\ref{fig:experiment_theory}), the odd mode (the left peak of the pair) with $m=11$ has a higher intensity than the even mode, while the odd mode with $m=$12 has a lower intensity and so on. 
It is interesting to compare the full sets of spectra measured along the axis in two types of nanotubes.  In the breathing NT (SI, Figure S6), the periodic variation in both intensity and mode energy ("splitting and merging") is observed, as predicted theoretically. In the flattened twisted nanotube with a nearly constant cross-sectional shape (SI, Figure S7), we observe  only a periodical change in the intensity of split modes, while the maxima of all the peaks remain relatively stable. This behavior confirms that in twisted nanotubes with a constant cross section, the effect of "splitting and merging" of the odd and even components of optical modes is impossible.


\section{Conclusion}

TMD twisted micro- and nanotubes demonstrate promising optical properties, such as the ability to emit light in a wide spectral range of 1.9-1.2 eV, along with the ability to support optical modes whose energy position can vary by $\sim$100 meV due to mode splitting in a strongly flattened cross section. This combination makes these tubes suitable for use as an efficient resonator for self-radiating light, with the ability to adjust the frequency at which the radiation will be amplified.

We have shown that intrinsic strain in the walls of nanotubes can lead to a modification of their cross-section shape, namely, to its flattening, accompanied by a rotation of the cross section along the tube axis. Such twisted NTs demonstrate exciton photoluminescence spectra modulated by the peaks of optical modes, which, in contrast to the whispering gallery modes in the cylindrical NTs, are split into even and odd components due the interaction of electromagnetic field in the opposite  walls in a strongly deformed cross section.

In PL spectra, the intensity of the split even and odd modes, being in antiphase, is very sensitive to the twist angle of the nanotube cross section with respect to the incident electric field.
We have developed a model describing at what aspect ratio of the cross section we will observe a deviation from the canonical WGMs and how changing the size of the gap in the breathing tubes with a variable cross section will affect the mode splitting energy. 
These findings suggest that applying an external force to such nanotubes can provide fine tuning of mode splitting, paving the way for the creation of new optomechanical devices.

Thus, the bright exciton photoluminescence, the high refractive index of the material, which keeps the optical modes inside the tube walls, and the flattened rotating cross section, which ensures mode splitting and a change in their intensity in antiphase, make twisted TMDC nanotubes a unique object. Owing to the observation of these effects at room temperature, one can predict their wide use in resonant nanophotonic structures and quantum emitters based on TMDCs.

\section{Methods}

\subsection{Micro-Raman and micro-PL measurements}

Micro-PL and micro-Raman measurements were carried out using a Horiba LabRAM HREvo UV-VIS-NIR-Open spectrometer equipped with a confocal microscope. The measurements were performed at room temperature with continuous-wave (cw) excitation using the 532 nm laser line of a Nd:YAG laser (Laser Quantum Torus). An Olympus MPLN100$\times$ objective lens (NA = 0.9) was used for both Raman and PL measurements, which allowed us to obtain information from an area with a diameter of $\sim1$ $\mu$m. To prevent damage to the structures, the incident laser power was limited to 1 mW.
PL spectra were obtained with a spectral resolution of 3.5 cm$^{-1}$ using a 600 gr/mm grating. The laser emission was linearly polarized, and a half-wave plate was used for its rotation. The Glan-Taylor prism was used 
as an analyzer. In case of Raman measurements, a spectral resolution of $\sim1.1$ cm$^{-1}$ was achieved using a 1800 gr/mm diffraction grating. To suppress the Rayleigh scattering and obtain information from the ultralow frequency range, a set of Bragg filters was used. 

\subsection{SEM characterization}

SEM characterization was performed using JEOL JSM 7001F scanning electron microscope. Details with extra images are presented in Supporting information.

\subsection{Financial Interest Statements}
The authors declare no competing financial interest.

\subsection{AUTHOR INFORMATION}
Corresponding Author

Ilya A. Eliseyev    

*E-mail: ilya.eliseyev@mail.ioffe.ru

ORCID

0000-0001-9980-6191

\subsection{Author Contributions}
The manuscript was written using contributions of all authors. In particular, I. A. E. and B. R. B - contributed equally to $\mu$-PL studies, and D. R. K. and A.V.P. - to theoretical modelling. 
All authors have given approval to the final version of the manuscript.

\begin{acknowledgement}
D. R. K., A. V. P., and T. V. S. thank the partial support within the Quantum Computing Roadmap (Agreements No. 868-1.3-15/15-2021 of 05.10.2021 and No. R2152 of 19.11.2021). D. R. K. and A. V. P. acknowledge the partial support from the grants \#SP-5068.2022.5  and \#MK-4191.2021.1.2, respectively. I. A. E. and V. Yu. D. acknowledge the partial support within the State Assignment \#0040-2019-0006.

\end{acknowledgement}

\bibliography{mos2_tubes}



\begin{suppinfo}

Procedure of the tube transfer, detailed SEM and TEM characterization results, optical image of the studied tubes, additional PL data.


\end{suppinfo}


\end{document}